\documentclass[aps,prc,groupedaddress,twocolumn,showpacs,amsmath,amssymb]{revtex4}
\usepackage{graphics}% Include figure files
\usepackage{dcolumn}% Align table columns on decimal point 
\usepackage{longtable}

\begin{document}

\title{Comment on ''Solution of the $\alpha$-Potential Mystery in the $\gamma$ Process and Its Impact on the Nd/Sm Ratio in Meteorites''}
\author{V.~Avrigeanu} \email{vlad.avrigeanu@nipne.ro}
\author{M.~Avrigeanu}
\author{C.~M\u an\u ailescu}
\affiliation{Horia Hulubei National Institute for Physics and Nuclear Engineering, P.O. Box MG-6, 077125 Bucharest-Magurele, Romania}

\begin{abstract}
A competition of the low-energy Coulomb excitation (CE) with the compound nucleus (CN) formation in $\alpha$-induced reactions below the Coulomb barrier has been assumed by Rauscher [Phys. Rev. Lett. {\bf 111}, 061104 (2013)] in order to use the same optical model (OM) potential for description of the latter and the $\alpha$-particle emission as well. In this Comment we show that the corresponding partial waves and integration radii provide evidence for the distinct account of the CE cross section and OM total-reaction cross section. 
\end{abstract}

\maketitle

\begin{figure*}  %[b]
\resizebox{2.06\columnwidth}{!}{\includegraphics{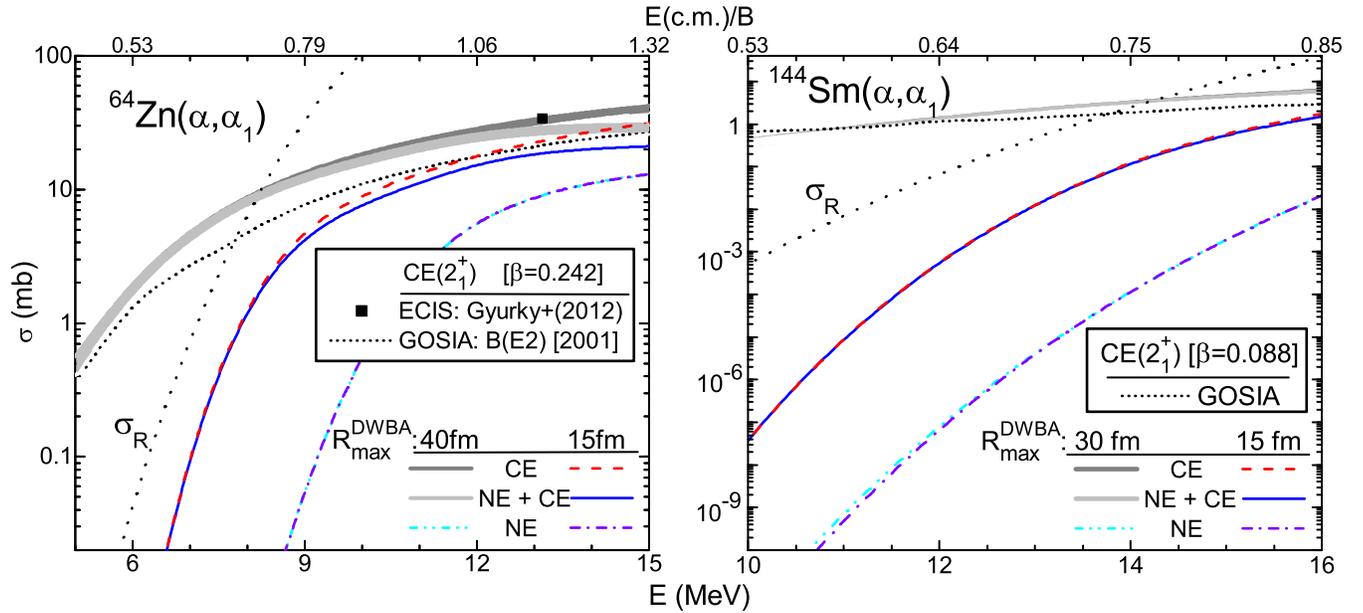}}
\caption{\label{Fig:Zn64Sm144CE} (Color online) Comparison of the $\alpha$-energy dependence of calculated total-reaction \cite{mcf66} (dotted curves) and CE cross sections for the first 2$^+$ excited state obtained by CC calculations \cite{gg12} (solid square), CE code GOSIA \cite{Gosia} (short-dotted curves), and code DWUCK4 using the radial integration up to either 15 (dashed curves) or 30-40 fm (dark shaded regions, see the text), for $\alpha$-particles incident on $^{64}$Zn (left) and  $^{144}$Sm (right) at energies around $B$. Cross sections for the collective inelastic scattering on the same states, for the same $R_{max}$ values, are also shown in the cases of NE+CE interference (solid curves and the light shaded regions, respectively), and NE alone (dash-dotted and dash-dot-dotted curves, respectively).}
\end{figure*}

{\it The "$\alpha$-potential mystery"} is referred by Rauscher \cite{tr13} to the competition of the Coulomb excitation (CE) with the compound nucleus (CN) formation in $\alpha$-induced reactions below the Coulomb barrier $B$, while the former does not affect the $\alpha$-particle emission. A diversion of the incident flux from the CN channel was thus taken into account on the basis of the CE semiclassical theory, leading to the decrease of the $\alpha$-particle transmission coefficients given for each partial wave by, e.g., the worldwide used optical model potential (OMP) of McFadden and Satchler \cite{mcf66}.  
Conversely, a distinct consideration of CE cross section and OM total-reaction cross section $\sigma_R$ is evidenced by the following cross-section assessment, and introduced elsewhere \cite{va16}.

Primarily, it should be noted that the same CE account has been found necessary \cite{tr13} to describe the measured $^{144}$Sm$(\alpha,\gamma)^{148}$Gd reaction cross sections \cite{es98} (including a further reduction by a factor of 3) and the $(\alpha,x)$ reactions for the target nuclei $^{141}$Pr \cite{as11} and $^{169}$Tm \cite{ggk11b}, but not for $^{130,132}$Ba \cite{zh12} and $^{168}$Yb \cite{ln13}. 
However, a reiteration of a pioneering study by Stelson and McGowan \cite{phs64} but using now the CE results of the semiclassical CC code GOSIA \cite{Gosia} and updated values of the corresponding reduced transition probabilities $B(E2)$ \cite{ensdf}, for the same $\alpha$-induced reactions \cite{tr13}, has shown that the CE cross sections for the target nuclei $^{130,132}$Ba and $^{168}$Yb have not remained small compared to the CN formation cross sections \cite{cm15}.

Moreover, a comparison of the OMP $\sigma_R$ \cite{mcf66} and CE cross sections provided by GOSIA is shown in Fig.~\ref{Fig:Zn64Sm144CE} for $\alpha$-particle inelastic scattering on the first 2$^+$ excited state of the target nuclei $^{64}$Zn and $^{144}$Sm, at energies around $B$. 
The fact that the CE cross sections for the first 2$^+$ excited state are several orders of magnitude larger than the $\sigma_R$ values given by the OMP analysis of the elastic scattering data, was shown actually much earlier \cite{phs64}.
However there are also shown in Fig.~\ref{Fig:Zn64Sm144CE} the results of distorted-wave Born approximation (DWBA) method calculations for the collective inelastic scattering on the same 2$^+$ excited states, carried out for the collective form factors corresponding to (i) pure either CE or nuclear excitation (NE), and their coherent interference (NE+CE) \cite{rhb62}, and (ii) integration radii $R_{max}$ of up to either 15 or 30 (40) fm which are typical to the short-range nuclear interactions (e.g., Ref. \cite{scat2}) and the long-range Coulomb field, respectively. The OMP \cite{mcf66} and the deformation-parameter values $\beta$ of 0.242 and 0.088 \cite{sr01}, respectively, have been used in this respect within the code DWUCK4 \cite{pdk84}. The radial integration up to 30 (40) fm has been considered in the latter case since the use of larger $R_{max}$ values provided no further increase of the CE cross sections. 
 
First, one may note that the pure NE cross sections are not increased by the extension of $R_{max}$, the two excitation functions corresponding to 15 and 30 (40) fm, respectively, being nearly overlapped. There are only 18 and 21 partial waves, respectively, contributing to these results at the $\alpha$-particle energies of 15 and 16 MeV, for the two target nuclei.  

Second, with the increase of the $\alpha$-particle energy and thus of the NE importance, the NE+CE cross sections are getting lower and lower with respect to pure CE. On the other hand, it should be underlined the same contribution of only 18 and 21 partial waves, respectively, at the upper energies of the excitation functions for the two targets.

Third, at the upper $R_{max}$ values for the two target nuclei shown in  Fig.~\ref{Fig:Zn64Sm144CE}, the CE as well as NE+CE cross sections are higher than OMP total-reaction cross section at $\alpha$-particle energies below $\sim$0.7$B$. The larger $R_{max}$ values correspond also to additional partial-wave contributions, the shaded regions concerning the increase from $\sim$20 to $\sim$60 partial waves contributing to both CE and NE+CE cross sections at the upper $\alpha$-particle energies in this figure. While an enlarged discussion on this issue was given earlier \cite{rhb62}, the use of either more partial waves or larger $R_{max}$ values has provided no further increase of the CE cross sections. 

It is thus proved that the largest contribution to CE and CE+NE cross sections comes from partial waves which are larger than the ones contributing to the OMP total-reaction cross section. It results that only the CE+NE cross sections of the $\alpha$-particle direct collective inelastic scattering that correspond to an integration radius which is typical to the short-range nuclear interactions, i.e., of $\sim$15 fm, should be subtracted from the OMP $\sigma_R$ in order to obtain the CN cross section which is to be then involved in statistical model calculations. 
Obviously, even below $B$, these CE+NE cross sections are much lower than the $\sigma_R$ values. 
Nevertheless, the question of the $\alpha$-potential mystery raised again by Rauscher \cite{tr13}, of largest interest for both nuclear astrophysics and nuclear technology \cite{ma06}, needs further consideration. 

{\it Acknowledgments.} This work was partly supported by Fusion for Energy (F4E-GRT-168-02), and by Autoritatea Nationala pentru Cercetare Stiintifica si Inovare (PN-42160102).

\end{document}